# THE GRAVISPHERE METHOD ALGORITHM PROGRAMMING

Rosaev A.E.[1]

[1] FGUP NPC "NEDRA" Svobody 8/38, Yaroslavl 150000, Russia, E-mail:rosaev@nedra.ru

ABSTRACTS: The action sphere method program is written. The initial conditions set at pericenter of planetocentric orbits. When action sphere radius is reached, the heliocentric orbit is calculated and data redirected to numeric integration program. The method is useful for capture and collision problem investigation. The very preliminary numeric results were obtained and discussed. A manifold in orbital elements space, leads to temporary capture about 50 year (4 Jupiter revolutions), was found.

## INTRODUCTION

When the motion takes place far from a planet (heliospheric region) the dominant effect is due to the Sun, while the dominant term of the equations of motion corresponds to the planet when the spacecraft moves in the so-called planetary gravispheres. The boundaries of the gravispheres can be defined in a number of different ways, such as: Laplace's spheres of influence, Hill's spheres [1], Belbruno's weak stability boundaries [2], etc.

According to this, the motion of the spacecraft can be expressed as a sequence of perturbed keplerian arcs. In the first order approximation, the trajectory is represented by a series of segments of undisturbed keplerian motion. Of course, matching conditions on both kinds of solutions must be added at the boundaries. The method is useful for capture and collision (launch) problem investigation. For the launch problem, the consideration of orbit of collision is required.

There are many practical tasks, leads to natural Solar system objects capture problem investigations, and many papers, concern with it. The most possible martian moons Phobos and Deimos origin is capture [3]. The Proto-Moon, before mega impact Earth, moved along orbit close Sun-Earth libration points [4] and probably, has a temporary capture on Earth satellite orbit before mega impact event. A number of irregular giant planet satellites placed on their modern orbit due to capture process. Jovian irregular satellites were studied in [5]. The capture in Neptunian system definitely took place [6,7]. The Uranian irregular satellites capture was modelled in [8]. In addition, capture mechanism may be important for asteroid hazard problem studying [9].

In a previous author's paper [10] some characteristics of capture orbit was derived. However, the real orbit of (temporary) capture was not constructed. Here we do it.

Gravity sphere radius determined by condition:

$$F_s = \frac{GM}{r^2} = F_{pl} = \frac{Gm}{R^2},$$

$$\rho = R_{p,s}\sqrt{\frac{m}{M}}$$

Here G-gravity constant, M, m – Sun and Jupiter mass respectively, R, r – heliocentric and planetocentric distance, $\rho$ – sphere radius, $R_{p,s}$ – target planet orbit radius.

Another gravispheres transformed respectively. Action sphere:

$$F_s/F_{pl} = F_{pl}/F_s,$$

$$\rho = R_{p,s}\sqrt[5]{\left(\frac{m}{M}\right)^2}$$

Somebody suggest, that for astrodynamical problem better:

$$\rho = 1.15\, R_{p,s}\sqrt[3]{\frac{m}{M}}$$

The expressions below provide transform from heliocentric to planetocentric orbital element and back at the gravisphere boundary.

$$x = r(\cos\Omega\cos(f+\omega) - \cos I \sin\Omega\sin(f+\omega))$$

$$y = r(\sin\Omega\cos(f+\omega) + \cos I \cos\Omega\sin(f+\omega))$$

$$z = r\sin I \sin(f+\omega)$$

$$r = \frac{a(1-e^2)}{1+e\cos f},$$

$$f = l + 2e\sin l + 5/4\, e^2 \sin 2l + ...,$$

Here $a$ – semimajor axis, $e$ – eccentricity, $I$ - inclination, $\omega$ – pericenter argument, $\Omega$ – accident node longitude, f – true anomaly, $l$ – longitude.

Velocities at elliptic motion are:

$$V_X = X/R \cdot V_R - (\sin(f+\omega)\cos(\Omega) + \cos(f+\omega)\sin(\Omega)\cos(I)) * V_N$$

$$V_Y = Y/R \cdot V_R - (\sin(f+\omega)\sin(\Omega) - \cos(f+\omega)\cos(\Omega)\cos(I)) * V_N$$

$$V_Z = Z/R \cdot V_R + (\cos(f+\omega)\sin(I)) * V_N$$

Here and below – x, y, z are rectangular coordinates of test particle in planetocentric frame, At this stage planet position and velocity entered:

$$X = X + X_0$$
$$Y = Y + Y_0$$
$$Z = Z + Z_0$$

$$VX = VX + Vx_0$$
$$VY = VY + Vy_0$$
$$VZ = VZ + Vz_0$$

$$V_E = \sqrt{V_{xE}^2 + V_{yE}^2 + V_{zE}^2}$$

$$R_E = \sqrt{x_{xE}^2 + y_{yE}^2 + z_{zE}^2}$$

New frame semimajor axis calculated from:

$$V = \sqrt{V_r^2 + V_n^2} = \sqrt{Gm\left(\frac{2}{r} - \frac{1}{a}\right)}$$

By taking into account:

$$(\mathbf{V*r}) = Vr \cos(V,r) = V_x * x + V_y * y + V_z * z$$

$$L = [V*R] = V_n * r = V_x * y - V_y * x$$

eccentricity directly calculated from:

$$L = V_n r = \sqrt{GMa(1-e^2)}$$
$$e = \sqrt{1 - (V_n r)^2 / (GMa)}$$

After that, the new frame inclination and node longitude calculated:

$$i = atg(\sqrt{(L_x^2 + L_y^2)} / L_z)$$

$$\Omega = atg(L_x / L_y)$$

Where:

$$L_X = Y \cdot V_Z - Z \cdot V_Y$$
$$L_Z = X \cdot V_Y - Y \cdot V_X$$
$$L_Y = X \cdot V_Z - Z \cdot V_X$$

At the transfer from heliocentric orbit, the described algorithm applied at two points: action sphere entry point and symmetry point.

Few very preliminary Results

The action sphere method program is written (fig.1). Here we report a very small example of our preliminary results. There are two series of calculations: with fixed semimajor axis (energy), and with fixed eccentricity of planetocentric orbit (table 1, 2). As a result, the good agreement with numeric integration is founded, but at early phase of encounter. The orbital elements continue to change.

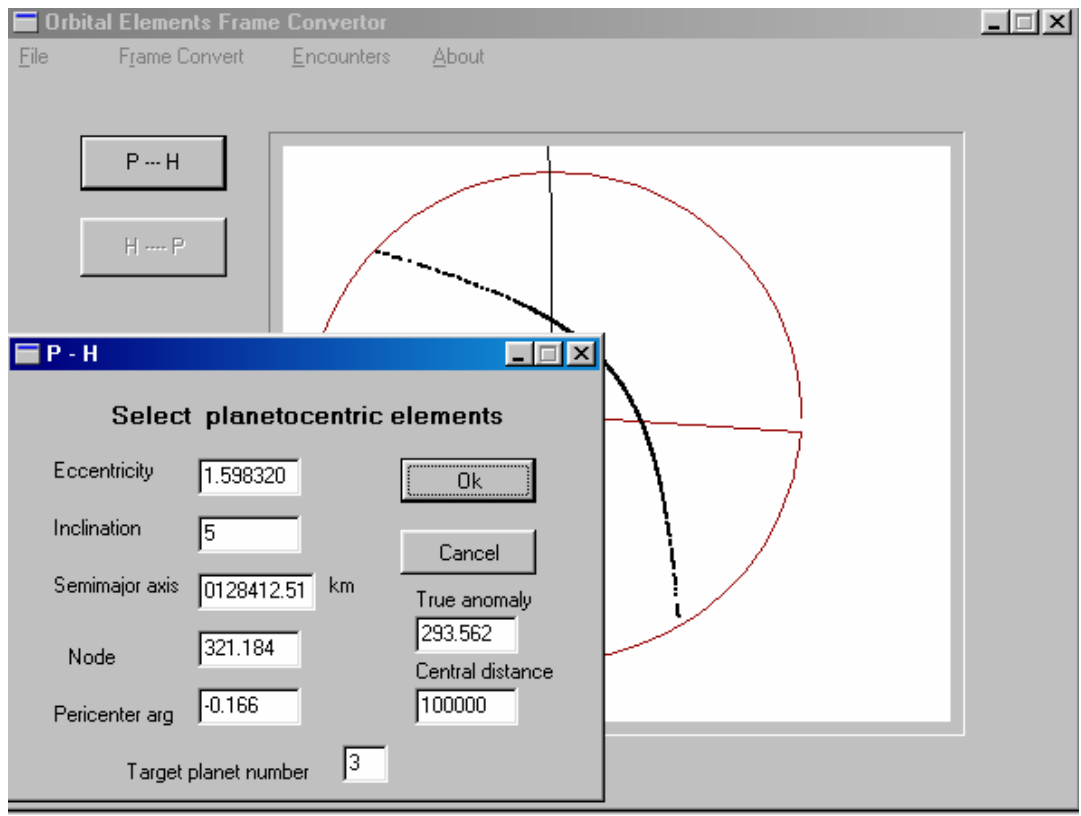

Fig.1. Program window

Table 1

Orbits with fixed energy

|   | Arrival orbit | Planetocentric orbit | Departure orbit |
|---|---|---|---|
| $a$ | 0.935479 | -0.16 mln. km | 1.005877 |
| $e$ | 0.094852 | 2.0 | 0.088075 |
| $a$ | 0.922136 | -0.16 mln. km | 1.021441 |
| $e$ | 0.104726 | 1.8 | 0.095962 |
| $a$ | 0.881170 | -0.16 mln. km | 1.092639 |
| $e$ | 0.131162 | 1.4 | 0.127002 |
| $a$ | 0.861945 | -0.16 mln. km | 1.142443 |
| $e$ | 0.146785 | 1.2 | 0.153929 |

Table 2

Orbits with fixed eccentricity

|   | Arrival orbit | Planetocentric orbit | Departure orbit |
|---|---|---|---|
| a | 0.905891 | 0.16 mln. km | 1.123891 |
| e | 0.085485 | 0.632 | 0.124752 |
| a | 0.902095 | 0.17 mln. km | 1.125545 |
| e | 0.089556 | 0.632 | 0.126410 |
| a | 0.900745 | 0.18 mln. km | 1.124823 |
| e | 0.090946 | 0.632 | 0.126121 |
| a | 0.896009 | 0.20 mln. km | 1.124358 |
| e | 0.096472 | 0.632 | 0.126870 |

The temporary capture orbit

The search for temporary capture on Jupiter satellite orbit was done. Jupiter orbit assumed circular. An initial conditions were set by varying the orbital elements in planetocentric frame. The fixed integration step was 876.58 seconds in main series of calculations. The calculation with step 87.658 seconds confirm fact of temporary capture.

A manifold in orbital elements space, leads to temporary capture about 50 year (4 Jupiter revolutions), was found (fig.2, 3). A very rough range of elements, leads to capture, is given in table 3. This manifold is close to circular orbit near to action sphere boundary.

Table 3

Range planetocentric orbit leads to Jupiter temporary satellite capture

|   | Range of elements, leads to capture |
|---|---|
| $a$, km | 22770000 – 27000000 |
| e | 0.01 – 0.04 |
| $\omega^o$ | 280 – 305 |
| $i^o$ | 0 – 8 |
| $\Omega^o$ | -12 – 15 |

Due to this manifold of orbit is continuum, it may be expected, that orbit of temporary capture of this class really present. Planetocentric orbit is very chaotic and has 'temporary loss' intervals. These intervals provide permanent loss of satellite at limit cases. The presence of temporary capture interval strongly depends on planetocentric semimajor axis. At $a$ < 22770000 km we have satellite-like motion, at $a$ > 27000000 km fly-by orbit. An example of heliocentric orbit before and after temporary capture phase is given in table 4.

Really, orbit before and after capture are the same orbit, rotated one from another.

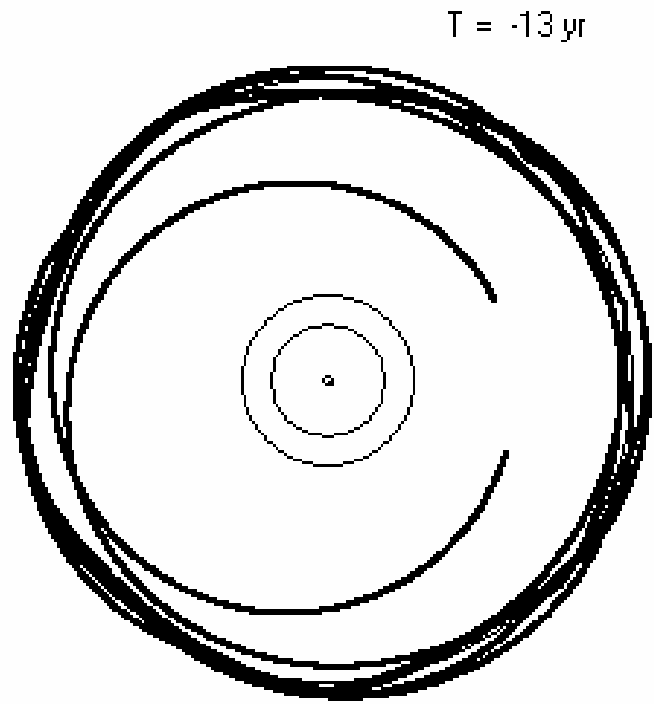

Fig 2. Heliocentric orbit of temporary Jupiter satellite capture

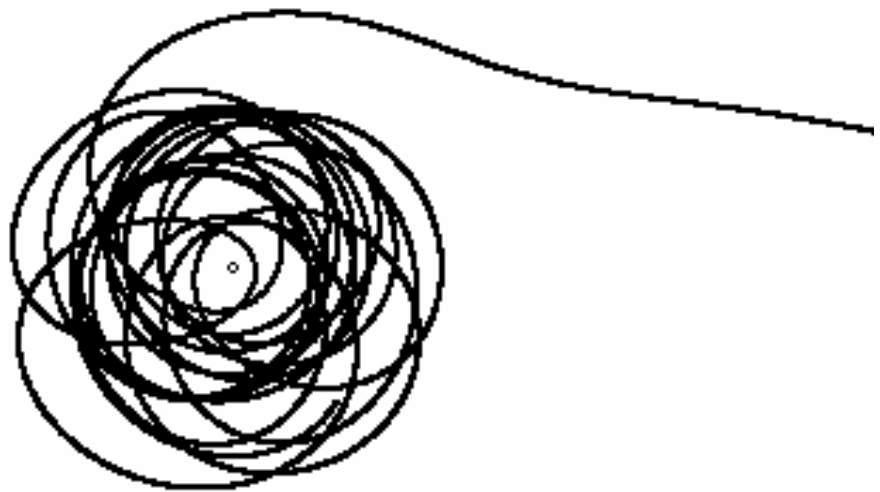

Fig.3. Planetocentric orbit of temporary Jupiter satellite capture

Table 4

Heliocentric orbit leads to Jupiter satellite capture

|  | Elements before capture | Elements after capture |
|---|---|---|
| $a$, a.u | 3.8955513 | 3.8893871 |
| e | 0.18052845 | 0.17267844 |

In addition, the search for temporary capture on Earth satellite orbit was done. Earth orbit assumed circular. A very chaotic orbit of temporary capture was found (table 5, fig.).

In addition, the series of calculations Earth temporary captured orbits was done. The results of Lidov [11] relative fast collision with Earth Moon, moved onto highly inclined orbit is confirmed (Fig. 4.)

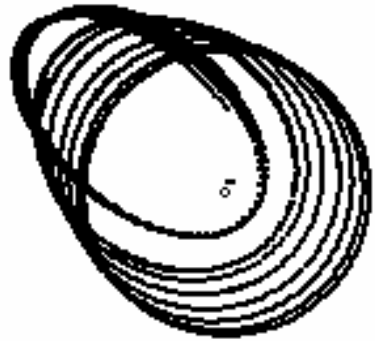

Fig.4. Moon at highly inclined (88) orbit collision with Earth within 5 years.

## Conclusions

The action sphere method program is written. The combination of action sphere method with numeric integration is very useful at capture/collision processes investigation.

The very preliminary numeric results were obtained and compared with numeric integration. By this way, a manifold in orbital elements space, leads to temporary capture about 50 year (4 Jupiter revolutions), was found. It is main result of this work.

# References

1. Spravochnoe rukovodstvo po nebesnoi mekhanike i astrodinamike. . /Eds G.N.Duboshin, 1976, Moscow, 864 p.
2. Belbruno E. 2004, Capture Dynamics and Chaotic Motions in Celestial Mechanics. Princeton, NJ : Princeton University Press, 211 p.
3. Singer S. F., 2003, Origin of Phobos and Deimos: a new capture model. *Sixth International Conference on Mars,* Abstr. No 3063.
4. Belbruno E., Gott J. R. Where Did The Moon Come From? arXiv:astro-ph/0405372 v2
5. Nesvorny D., Beauge C., Dones L. 2004, Collisional Origin of Families of Irregular Satellites *The Astronomical Journal*, **127**: 1768–1783.
6. Goldreich, P., Murray, N., Longaretti, P. Y. & Banfield, D. Neptune's story. Science 245, 500–-504 (1989).
7. Cuk M. and Gladman B. J. Constraints on the Orbital Evolution of Triton. arXiv:astro-ph/0505235 v1.
8. Vieira Neto E., Winter O. C. 2001, Time Analysis for Temporary Gravitational Capture: Satellites Of Uranus. *The Astronomical Journal*, 122, 440-448.
9. Rosaev A.E. 2006, Studying Close Encounters With Gravisphere Method: International Astronomical Union. Symposium no. 236, held 14-18 August, 2006 in Prague, Czech Republic, S236, #58
10. Rosaev A.E. 2003, The one kind of orbit of collision, related with Lagrangian libration points. Proceedings of Libration Point Orbits And Applications Conference, Eds. by  G. Gomez, M. W. Lo and .J.J. Masdemont, World Scientific Pub, New Jersey - London – Singapore - Hon Kong. p. 613-622.
11. Lidov M.L. On the approximate analysis of evolutions of orbits artificial satellites.// Problems of moving artificial celestial bodies. M., Nauka, 1963, P. 119-134.